\documentclass{mem}
\usepackage{natbib}\usepackage{txfonts}\usepackage{balance}
\usepackage{graphicx}
\usepackage{txfonts}
\usepackage[a4paper]{hyperref}
\idline{79}{3}
\newcommand{\strom}{\mbox{Str\"omgren~}}

\begin{document}
\def\teff{$T\rm_{eff }$}
\def\kms{$\mathrm {km s}^{-1}$}

\title{\strom metallicity calibration: the $m_1,\ b-y$ relation}

   \subtitle{}

\author{
A. \,Calamida\inst{1} 
\and  G. \, Bono\inst{1}
\and  P. B. \,Stetson\inst{2}
\and  L. M. Freyhammer\inst{3}
\and  S. \, Cassisi\inst{4}
\and  F. \, Grundahl\inst{5}
\and  A. \, Pietrinferni\inst{4}
\and  M. \, Hilker\inst{6}
\and  F. \, Primas\inst{6}
\and  T. \, Richtler\inst{7}
\and  M. \, Romaniello\inst{6}
\and  R. \, Buonanno\inst{8}
\and  F. \, Caputo\inst{1}
\and  M. \, Castellani\inst{1}
\and  C. E. \, Corsi\inst{1}
\and  I. \, Ferraro\inst{1}
\and  G. \, Iannicola\inst{1}
\and  L. \, Pulone\inst{1}
          }


\institute{
INAF-OAR, via Frascati 33, Monte Porzio Catone, Italy
\email{calamida@mporzio.astro.it}
\and
DAO, HIA, National Research Council, Victoria, BC, Canada
\and
University of Central Lancashire, UK
\and
INAF-OACTe, Teramo, Italy
\and
Department of Physics and Astronomy, Aarhus University, Denmark
\and
ESO, Garching, Germany
\and
Universidad de Concepcion, Concepcion, Chile
\and
Universit\'a di Roma Tor Vergata, Rome, Italy
}

\authorrunning{Calamida et al.}

\titlerunning{\strom metallicity calibrations}

\abstract{We performed a new calibration of the \strom metallicity index 
$m_1$ based on the $b-y$ color of cluster red giant stars. The current 
Metallicity-Index-Color (MIC) relation is not linear in the color range
$0.40 \lesssim b-y \lesssim 1.0$, but provides iron abundances 
of cluster and field red giants with an accuracy of $\sim$ 0.25 dex.
\keywords{globular clusters: general --- stars: abundances --- stars: evolution}
}
\maketitle{}

\vspace{-0.2truecm}
\section{Introduction}
Empirical calibrations of the \strom metallicity index 
$m_1$=$(v-b)-(b-y)$ were usually based either on field stars 
(Anthony-Twarog \& Twarog 1998, hereafter ATT98) or on a mix of 
cluster and field stars (Schuster \& Nissen 1989; Hilker 2000).  
However, empirical spectroscopic evidence suggest that 
field and cluster stars present different heavy element abundance
patterns (Gratton, Sneden \& Carretta 2004). Moreover, the occurrence 
of CN and/or CH rich stars in globular clusters (GCs, Grundahl, Stetson, \& Andersen 2002)
along the red giant (RG, Hilker 2000) and the subgiant branches, and the main
sequence (Kayser et al. 2006) opens the opportunity for an independent 
calibration of the \strom metallicity index only based on cluster stars.

Calamida et al. (2007a, hereafter CA07) provided new empirical
calibrations of the $m_1$ metallicity index based only on cluster RG stars.
The new Metallicity-Index-Color (MIC) relations have been validated by adopting 
GC and field RGs with known spectroscopic iron abundances in the Zinn \& West (1984) 
scale, and provide metallicity estimates with an accuracy $\le 0.2$ dex. 
The main advantage of CA07 MIC relations compared to similar relations 
available in the literature is that they adopt the $u-y$ and $v-y$ colors 
instead of the $b-y$. The main advantage of the $u-y$ and $v-y$ colors
is the stronger temperature sensitivity, and 
the MIC relations show a linear and well-defined slope in the $m_1,\ u-y / v-y$
planes (see CA07).  
We now perform a new empirical metallicity calibration 
based on the $b-y$ color, and show that the $m_1,\ b-y$ relation for cluster 
RGs is not linear over the color range $0.42<(b-y)_0<1.05$. We then 
compare the new metallicity calibration with the CA07 $m_1,\ v-y$ MIC relation. 

\vspace{-0.35truecm}
\section{Metallicity estimates}
In order to calibrate the metallicity index $m_1$ we selected four GCs, 
namely M92, M13, NGC~1851, NGC~104, that cover a broad range in 
metallicity ($-2.2<[Fe/H]<-0.7$), are marginally affected by reddening 
($E(B-V) \le 0.04$), and for which accurate $u,v,b,y$ \strom photometry 
well-below the Turn-Off region is available. The reader interested
in details concerning the observations, data reduction and calibration procedures is referred 
to CA07\footnote{The \strom catalogs adopted in this investigation can be retrieved 
from the following URL: 
{\scriptsize http://www.mporzio.astro.it/spress/stroemgren.php}}. 
We selected stars from the tip to the base of the RGB for each 
cluster in our sample. However, in order to avoid subtle systematic uncertainties 
in the empirical calibration, current cluster RG stars have been cleaned for the
contamination of field stars. To accomplish this goal we used optical-Near-Infrared
(NIR) color planes to split cluster and field stars. All cluster samples were
reduced by $\approx$ 40\% after the color-color plane selection.
We derived a MIC empirical relation that correlate the iron abundance of RG 
stars to their metallicity ($m_1$) and $b-y$ color. We assumed 
$E(b-y)= 0.70\times E(B-V)$, adopting the reddening law from Cardelli et al. (1989) 
and $R_V=3.1$. Reddening values for the selected clusters are from Harris (2003) and 
Schlegel et al. (1998). 
We applied a multilinear regression fit, by adopting 
cluster metallicities, to estimate the coefficients of 
the MIC relation. Note that the $m_1,\ b-y$ relation is not linear in the
selected color range ($0.42 <(b-y)_0< 1.05$) and we had to include a quadratic 
color term. The metallicities adopted in the fit are in the Zinn \& West scale.

In order to validate the new empirical calibration we decided to apply it 
to five GCs for which accurate \strom photometry, absolute calibration,
and sizable samples of RG stars are available. They are NGC~288, NGC~362, 
NGC~6752, NGC~6397, and M71.
\begin{figure}[ht!]
\centering
\begin{minipage}[l]{.45\textwidth}
\centering
\includegraphics[height=5.cm,width=5.cm]{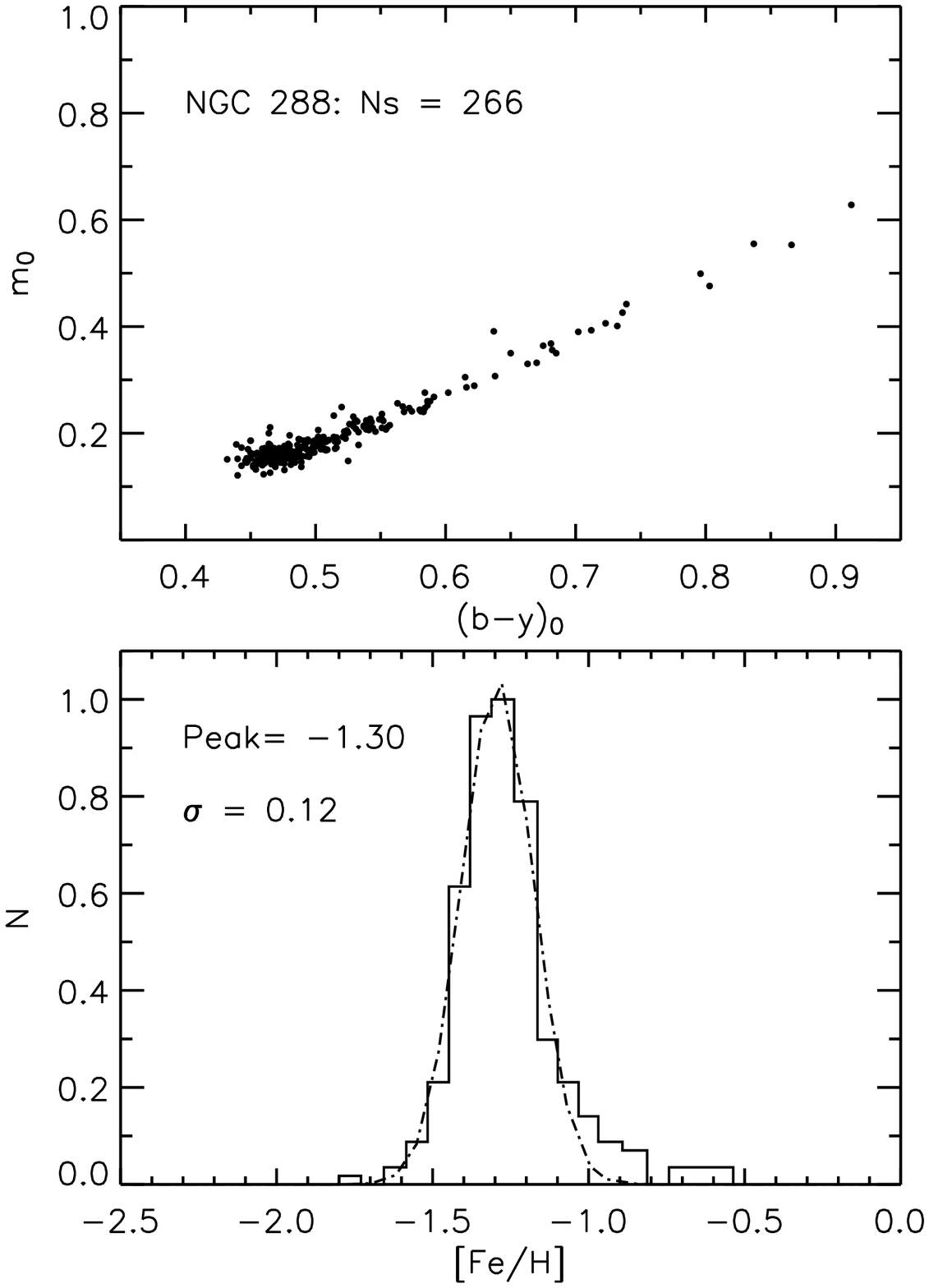}
\vspace{-0.3truecm}
\caption{\footnotesize{Top: RGs in NGC~288 plotted in the ($m_{0},\ (b-y)_0)$ plane. 
Candidate cluster stars were selected according to the optical-NIR color plane. 
Bottom: Metallicity distribution obtained applying the 
$m_{10}, (b-y)_0$ MIC relation to the RGs of the top panel. 
The distribution was fitted with a Gaussian (dashed-dotted line). The peak 
value and the dispersion are labeled.} \label{fig:fig1}}
\end{minipage}
\begin{minipage}[r]{.45\textwidth}
\centering
\includegraphics[height=7.5cm,width=6cm]{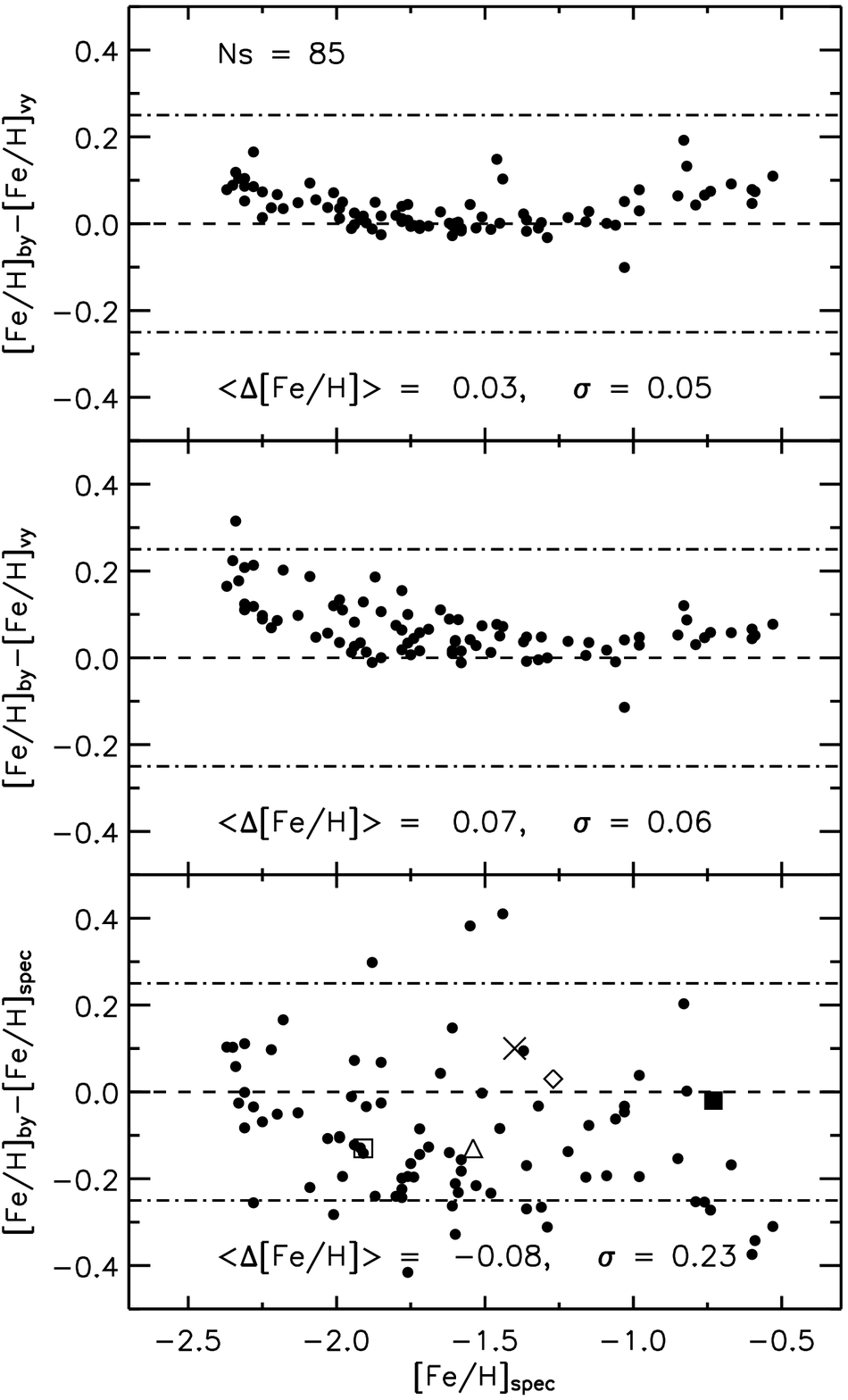}
\vspace{-0.27truecm}
\caption{\footnotesize{Difference between photometric $[Fe/H]_{vy}$ 
and $[Fe/H]_{by}$ metallicities for the 85 field RGs by ATT98. The $[Fe/H]_{by}$
metallicities are estimated adopting a quadratic (top panel) and a linear (middle)
$b-y$ color term. Bottom: Difference between photometric ($[Fe/H]_{by}$, quadratic) 
and spectroscopic metallicities plotted versus $[Fe/H]_{spec}$ for the 85 RGs.
Different symbols show the metallicity estimates of the five validation GCs: 
-empty square: NGC~6397, -empty triangle: NGC~6752, - cross: NGC~288, 
-empty diamond: NGC~362, -filled square: M71.}\label{fig:fig2}}
\end{minipage}
\end{figure}
Fig.~1 (top panel) shows NGC~288 RG stars plotted in the $m_{0},\ (b-y)_0$ plane after the
optical-NIR color--color selection. The bottom panel of Fig.~1 shows the metallicity 
distribution obtained for the RGs using the new MIC relation between $m_0$ and $(b-y)_0$.
The  distribution was fitted with a Gaussian function, 
with a peak value of $[Fe/H]_{phot} = -1.30$ dex and a dispersion of 
$\sigma = 0.12$ dex (dashed-dotted line). This cluster metallicity estimate 
agrees quite well with the spectroscopic result ($[Fe/H]_{spec} = -1.40\pm0.12$, see
Table~1). 
We find a similar agreement also for the other four GCs we adopted to validate 
the MIC relation.  Fig.~2 shows the comparison between the mean photometric 
metallicity estimates (listed in Table~1) and the spectroscopic measurements 
(bottom panel). Data plotted in this figure indicate that they agree with each other within 
$1\sigma$ errors. Together with the validation based on GC stars we decided to 
test the accuracy of the new MIC relation using a sample of field RG stars. 
\strom photometry, spectroscopic abundances, and reddening estimates for these 
stars were retrieved by Anthony-Twarog \& Twarog (1994; 1998, CA07 and references therein). 
We selected only RGs in the metallicity range $-2.4<[Fe/H]<-0.5$ and in the color range 
$0.42 <(b-y)_0< 1.05$. Fig.~2 shows the comparison between CA07 $m_1,\ v-y$ MIC relation 
and current $m_1,\ b-y$ relation (top panel) and a linear $m_1,\ b-y$ relation (middle
panel). It is clear that the parabolic trend between the photometric $[Fe/H]_{vy}$ 
and $[Fe/H]_{by}$ metallicities decreases when adopting a quadratic $b-y$ color term 
in the MIC relation.
The bottom panel of Fig.~2 shows the difference between the photometric and
spectroscopic metallicities for the 85 field RGs as a function of their 
spectroscopic iron abundances. This figure shows that on average 
there is a systematic shift of $\sim-0.1$ dex towards more metal-poor values, 
as already found for the $m_1,\ v-y$ MIC relation by CA07.
In spite of this systematic difference and the mild residual parabolic trend, 
the intrinsic dispersion is smaller than 0.25 dex.


\vspace{-0.35truecm}
\section{Conclusions}
We present a new empirical calibration for the \strom $m_1$ metallicity index
based on the $b-y$ color. We included a quadratic color term in the MIC relation in
order to account for the non-linearity of the $m_1,\ b-y$ relation for RG stars in the
color range $0.42 < (b-y)_0 < 1.05$. The new MIC relation provides metallicity estimates
for both cluster and field RGs with an accuracy of $\sim$ 0.25 dex.

\begin{table}
\caption{\footnotesize{Spectroscopic measurements from Rutledge et al. (1997) 
and photometric metallicity estimates for the GCs adopted to validate the 
\strom metallicity index.}}
\label{table1}
\begin{scriptsize}
\begin{center}
\vspace{-0.55truecm}
\begin{tabular}{lcc}
\hline
\\
GC  &  $[Fe/H]_{spec}$ & $[Fe/H]_{phot}$ \\
\hline
\\
NGC~6397 & $-1.91\pm0.14$ & $-2.04\pm0.15$ \\
NGC~6752 & $-1.54\pm0.09$ & $-1.67\pm0.18$ \\
NGC~288  & $-1.40\pm0.12$ & $-1.30\pm0.11$ \\
NGC~362  & $-1.27\pm0.07$ & $-1.24\pm0.30$ \\
M71	 & $-0.73\pm0.05$ & $-0.48\pm0.34$ \\
\\
\hline
\end{tabular}
\end{center}
\end{scriptsize}
\end{table}


\vspace{-0.35truecm}
\bibliographystyle{aa}

\end{document}